# Feature Engineering on LMS Data to Optimize Student Performance Prediction

Keith Hubbard[1][0000-0002-7002-1697] and Sheilla Amponsah[1][0009-0000-9019-5064]

[1] Stephen F. Austin State University, Nacogdoches, TX, USA

**Abstract.** Nearly every educational institution uses a learning management system (LMS), often producing terabytes of data generated by thousands of people. We examine LMS grade and login data from a regional comprehensive university, specifically documenting key considerations for engineering features from these data when trying to predict student performance. We specifically document changes to LMS data patterns since Covid-19, which are critical for data scientists to account for when using historic data. We compare numerous engineered features and approaches to utilizing those features for machine learning. We finish with a summary of the implications of including these features into more comprehensive student performance models.

## 1      Introduction

College student success and attrition, along with their connection to student engagement have been a frequent topic of research analysis (Kuh et al., 2008; Kuh et al., 2010; Hutt et al., 2018; Macnamara & Burgoyne, 2023). Employing data science on these topics is logical since the volume of data is so large. There are classroom-level learning analytics approaches focusing on a single context as well as student and faculty experiences with the data analysis (eg. Arnold & Pistilli, 2012; Dietz-Uhler & Hurn, 2013). There has been work on unpacking the data within learning management systems (LMS) to support student success (eg. Broadbent, 2016). However, Marques and colleagues' analysis concluded that, in their context, LMS data alone was insufficient to meaningfully predict student success (2017). Early alert systems are now used extensively across higher education as summarized in Velasco's metanalysis of the systems. Her research concludes that results vary with timing, staff training, and a variety of other factors, implying that the details of an institution's data utilization and interventions matter far more than just 'having an alert system' (2020).

In another approach to leveraging student data, Wong attributed a 4% increase in first-year retention at least partially to utilizing students' incoming data, orientation attendance, survey responses, and midsemester grades in a dashboard available to staff (2021). Other aspects of college students' data footprint have also been studied. College students' meal plan usage has been used to model student social networks (Bowman et al., 2019). Another study highlights the potential for meal usage data to support interventions for student success *in theory*, but were not able to produce significant predictors (Samuel & Scott, 2014). DeCarbo credits predictive analytics with improving course outcomes at one institution (without detail on methodology) but also concluded only a fifth of higher education institutions are identifying students who are at risk academically on any wide-spread level (2022). Pistilli's analysis shows how hard it is to utilize most predictive data on a timeline that maximizes its utility (2017). It seems that the underlying theme is that although extensive data is being gathered on students, optimizing its use requires attention to the specifics of that data.

Crossley et al. examined MOOC data for 320 students in a 2013 course called Big Data in Education. They analyzed click-stream data and natural language processing. They were able to predict with 78% accuracy which students would successfully complete the course.

Bird and colleagues examined a state-wide community college system attempting to create models to predict whether a particular student would pass a particular course. They compared using administrative data with using LMS data, and also a combination of the two. They also considered new students separately from returning students, finding that the LMS data was most helpful in improving predictions for new students, but minimally helpful for returning students (in press). This was a large study incorporating 226,784 students. Data collection ended in Spring 2021 (when only 75% of their courses contained any LMS activity), so the study is recent, but our research points to changes in LMS usage even since then.

The rapid growth of LMS's, and online learning in general, is well documented. An estimated 98% of universities moved online in April 2020 due to the covid-19 pandemic (Sadler, 2021). Although many institutions largely returned to face-to-face education, particularly in 2022, Bouchrika documents how LMS usage is again growing rapidly at universities (as high as 17% annually in Asia) (2025).

This article has three focuses:

- Examine how LMS Login and Grade data have changed over recent years at our institution, which has implications for training models on historic data;
- Examine how feature engineering affects the predictive power of LMS Login and Grade data – arguably the two most important features within LMS's;
- Examine how LMS data improves student performance models utilizing general student data.

## 2      Method

The research group from which this work emerges has grade data on all undergraduate students at a midsized regional comprehensive university dating back to 2010. We also have comprehensive LMS data dating back to 2016, however the nature of this data changes dramatically over time. We begin our methodology by explaining the rationale for the time periods on which our study focused. We then move to the rationale for the LMS features we will focus on in this work.

### 2.1 Timeframe Selection

Recall that Spring 2020 was the term when Covid-19 led to the reformatting of college courses worldwide. Our institution was already regularly utilizing an LMS system prior to 2020, so the change might not be as drastic as other institutions, but Table 1 outlines the change in LMS "Logins per Student" at the time of the Covid-19 outbreak. We use this table as a rationale to limit our attention to Fall 2020 through Fall 2024 data, since on some level the logins per student appear to stabilize over that period.

**Table 1.** Students enrolled full-time (12+ hours) and LMS logins of those students

|             | Total Students | Total Logins | Logins per Student |
|-------------|---------------:|-------------:|-------------------:|
| Spring 2019 | 8,318          | 1,668,586    | 201                |
| Fall 2019   | 9,140          | 1,916,127    | 210                |
| Spring 2020 | 8,092          | 1,874,798    | 232                |
| Fall 2020   | 8,809          | 2,252,395    | 256                |
| Spring 2021 | 7,669          | 1,923,893    | 251                |
| Fall 2021   | 8,236          | 1,932,580    | 235                |
| Spring 2022 | 7,030          | 1,726,979    | 246                |
| Fall 2022   | 7,764          | 1,655,993    | 213                |
| Spring 2023 | 6,675          | 1,348,794    | 202                |
| Fall 2023   | 7,735          | 1,861,928    | 241                |
| Spring 2024 | 6,686          | 1,815,653    | 272                |
| Fall 2024   | 8,101          | 2,171,345    | 268                |
| **Total**   | **94,255**     | **22,149,071** | **235**          |

The table also restricts attention to "long-semesters," meaning no summer terms. We chose to restrict to these terms since course structures differ dramatically at our university over the summer (with much smaller numbers of students).

We first examined registered logins for the Brightspace Desire2Learn LMS over nine long semesters from Fall 2020 to Fall 2024. There were a total of 19,977,140 logins by 27,875 unique users over the period of consideration. We associated this data with information about the users. We then disaggregated these logins by term, by major, by gap between login, and by other measures attempting to understand which aspects of logins yielded the best predictive power for students' semester grade point average (GPA), their overall GPA, and their retention at the university.

We also examined all grades assigned within the LMS over the same period. There were a total of 10,129,981 grades assigned to a total of 29,233 distinct students. Again, grades were associated with underlying student data, examined and disaggregated. Ultimately, we attempted to utilize these grades to optimize prediction of student GPA and retention at the university.

In total we considered 85,848 undergraduate student-semester outcomes from the nine semesters of interest. For each of these student-semesters, it would be useful to have predictions as to how the student performed (GPA) and whether the student was retained. A total of 23,471 distinct courses are represented.

### 2.2   LMS Feature Selection

The most dominant features in predicting student performance are prior student GPA and midsemester grades (which are provided by faculty to indicate what grades are likely to be at the end of the term). These features are generally widely available to university personnel and fairly simple to interpret.

In contrast, LMS data is *not* widely available to university personnel and not well understood in the aggregate. Specifically, at our university neither a student's academic advisor nor any of their individual professors has access to *all* of their LMS grades, discussion posts, logins, or any other features. By design, Learning *Management* Systems are supposed to track or manage the learning / performance a student does, so there is every reason to believe this data should be profoundly predictive of a student's academic performance. Our team examined 72 features from the LMS and settled on around 10 features that showed particular promise. These features were cleaned, missing data was analyzed, and the feasibility of resultant engineered features was analyzed. We delay detailed descriptions of how features were engineered until the Analysis section of this paper, but examine feature importance, first in terms of correlation to three outcomes of interest: Students' *semester GPA*, Students' *overall GPA* at the end of semester, and Students' *discontinuance* (or whether they left the university at the end of the semester without a degree). Table 2 presents the top 10 features correlated to each of the three outcomes of interest.

**Table 2.** Top 10 features, ranked by correlation to 3 key outcomes.

| Semester GPA | Overall GPA | Discontinuance |
| --- | --- | --- |
| Midterm grades | Beginning GPA | High school |
| ***LMS Grades*** | Midterm grades | ***LMS Periodic Logins*** |
| Beginning GPA | ***LMS Grades*** | Student type code |
| Alerts | High school | ***LMS Grades*** |
| High school | High school percentile | Major |
| ***LMS Periodic Logins*** | Alerts | Total hours attempted |
| High school percentile | Major | TSI score |
| Major | LMS content completion | Midterm grades |
| LMS content completion | ***LMS Periodic Logins*** | Beginning GPA |
| LMS content visits | LMS content topics | Admit code |

Observe that "LMS grades" appears as the second, third, and fourth most significant predictors in these lists, while LMS Periodic Logins appears as the sixth, nineth, and second most significant logins. Other LMS features appear, but never ranking higher than seventh.

We also employed Recursive Feature Elimination (RFE) to a random forest model for each of the three outcomes of interest. We used a random forest model initially because nearly all models which have historically performed best on this type of data are aggregations of decision trees. Table 3 summarized the top five features on the three outcomes of interest.

**Table 3.** Top 5 features, ranked by RFE on the 3 key outcomes

| Semester GPA | Overall GPA | Discontinuance |
| --- | --- | --- |
| Midterm grades | Beginning GPA | High school |
| ***LMS Grades*** | Midterm grades | ***LMS Grades*** |
| Beginning GPA | ***LMS Grades*** | Total hours attempted |
| High school | High school | Midterm grades |
| ***LMS Periodic Logins*** | Total hours complete | Beginning GPA |

Observe that the only two LMS features to appear in Table 3 after RFE are LMS Grades and LMS Periodic Logins. With these justifications, we move onto our analysis of features and feature engineering pertaining the LMS grades and logins.

## 3  Analysis

We consider LMS login data, then LMS grade data. Finally, we summarize how these data combine with other student data for more comprehensive prediction models.

### 3.1  LMS Login Data

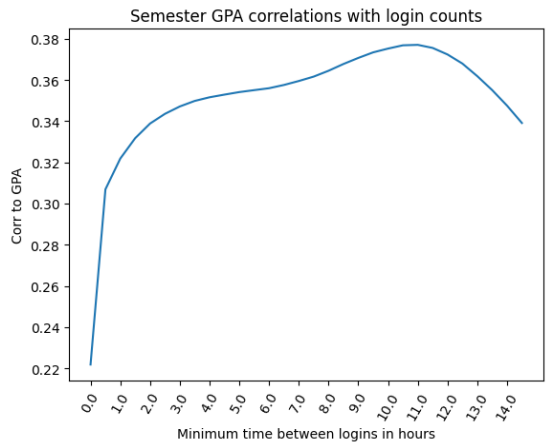

**Figure 1.** Correlation between semester GPA and login counts, only counting logins with the specified minimum login gaps

Calculating a correlation coefficient between the number of student logins during a term and their semester GPA yields 0.22. However, many logins are quite close together. By restricting our consideration to logins at least a minimum distance apart, we are able to improve the relationship to semester GPA as Figure 1 illustrates. In fact, the highest correlation coefficient is 0.38 when removing all logins less than 11 hours apart.

We also examined ignoring logins that were spaced beyond a given maximum. This produced marginal correlation gains but did not combine effectively with the substantial gains seen by imposing minimum login gaps. Additionally, we examined counting 24-hour or 12-hour periods with a login for each student. Neither performed as well as the minimum login gaps illustrated in Figure 1.

As seen in Figure 2, the predictive power of logins varies dramatically between different major types with the highest being 0.55 correlation for Psychology majors (peaking at 11 hours gaps) and the lowest being 0.20 correlation for Dual Credit students (peaking at 10 hours gaps). Note that all peak correlations occur between 10 and 11 hours, indicating that this feature is robust across subpopulations.

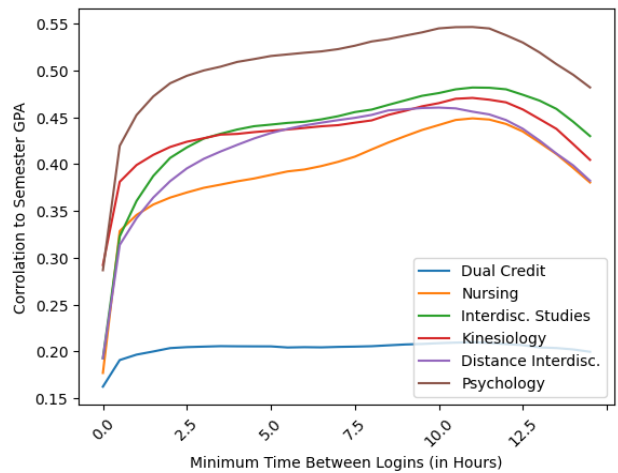

**Figure 2.** Correlations between semester GPA and login counts, with specified minimum login spacings, disaggregated by the top 6 majors

We also examine the correlations of logins with credit hours *attempted* and credit hours *completed* as illustrated on the left and the right of Figure 3, respectively. Students who attempted 15 or more credit hours had the highest correlation at 0.49 (peaking at 10.5 hours gaps) and the lowest was 0.22 for students with at most 3 credit hours attempted (peaking at 11 hours gaps). For credit hours completed, students having completed between 31 to 60 credit hours (peaking at 10.5 hours gaps) had the highest correlation between periodic logins and semester GPA at 0.5. Tied for lowest correlation, 0.3, were students with at most 30 total credit or more than 90 total credit hours. Note that all peak correlations occur between 10.5 and 11.5 hours, indicating that this feature is again robust across subpopulations.

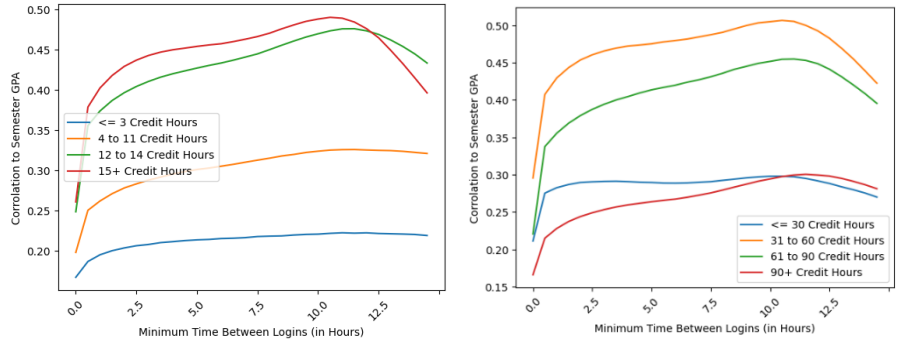

**Figure 3.** Correlations between semester GPA and login counts, with specified minimum login spacings, disaggregated by the credit hours attempted on the left and credit hours completed on the right

We focused our efforts on interpreting the utility of *periodic* login counts with a minimum 11-hour gaps between logins. Deploying flaml to test a variety of regressors to use this single variable to predict students' semester GPA, XGBoost regressor performed best. It produced a prediction on the test set (which was 20% throughout) an MSE of 0.818; 42% of the test set GPAs were predicted within 0.5 grade point; and 77% of the test set were predicted within 1 grade point.

Since predictive power is more important early in the semester for students, we also restricted to periodic logins in the first 4 weeks, first 8 weeks, and first 12 weeks (not counting Spring Break), training models on each. We also built one model using all four timeframes to allow for the possibility that a *change* in the number of logins throughout the semester was a significant feature. As seen in Table 4, the models were not wildly disparate. Notice that all of the models are decisions tree aggregators and four-of-five employ gradient boosting. Models were compared on 1- $R^2$. For performance comparison, on the "All Combined" modeling LMGM did best at 0.7937, then RF at 0.7955, then XGBoost at 0.8044, then ExtraTrees at 0.982, then XGB_LimitDepth at 0.9068, then SGD at 0.9585.

**Table 4.** The performance of periodic login data from different portions of the semester in predicting semester GPA

|  | First 4 Weeks | First 8 Weeks | First 12 Weeks | Full Semester | All Combined |
|---|---|---|---|---|---|
| Periodic logins | 1,849,465 | 3,475,366 | 5,173,306 | 6,705,135 | n/a |
| MSE | 0.899 | 0.882 | 0.874 | 0.818 | 0.760 |
| within 0.5 point | 40% | 41% | 41% | 42% | 43% |
| within 1 point | 76% | 77% | 77% | 77% | 79% |
| Best model | LGBM | LGBM | ExtraTrees | XGB | LGBM |

Login data was less effective in predicting discontinuance. Again utilizing flaml, we examined classification models over the same time frames as documented in Table 5.

8       K. Hubbard and S. AmponsahThese is some improved prediction power with more weeks of data, but none of the numbers are impressive. Again, models performed moderately similarly (comparison using area 1 – "area under ROC"). On the "All Combined" model, XGBoost performed best (0.2498), then RF (0.2510), then LGBM (0.2523), then XGB_LimitDepth (0.2565), ExtraTrees (0.2566), then SGD (0.3446).

**Table 5.** The performance of periodic login data from different portions of the semester in predicting discontinuance

|  | First 4 Weeks | First 8 Weeks | First 12 Weeks | Full Semester | All Combined |
|---|---|---|---|---|---|
| Accuracy | 0.89 | 0.89 | 0.89 | 0.89 | 0.89 |
| Area under ROC | 0.67 | 0.70 | 0.73 | 0.74 | 0.76 |
| Best model | ExtraTrees | ExtraTrees | RF | LGBM | XGB |

Note that in the context of identifying and connecting with students who are discontinuing, the overall ROC curve is less critical than the endpoints. Judged through a student retention lens, for expensive student interventions one requires high precision and for inexpensive / bulk interventions really only a high recall is required.

### 3.2   LMS Grade Data

We now turn our attention to grade data within the LMS. Instructors have the ability to add every grade, numeric or not, into the LMS to allow students to monitor their progress. Most universities do not have a requirement that all, or even any, course grades be entered into an LMS. The university under consideration is no different.

Table 6 summarizes number of students, courses, and LMS grades across terms. Observe that student count is down slightly over times, and grade volume is up. Combined, we see that the grades-per-student average over the first two terms in the dataset is 90, while in the last two terms an average of 136 grades-per-student are recorded, a 51% increase. This increase resulted almost entirely from grades that were updated before the end of the semester. The increase in usage likely came first from the push to LMS communication in the immediate aftermath of Covid-19, but then moved earlier in the term as faculty more robustly incorporated LMS grading into their courses rather than just posting grades toward the end of the term.

We retained all copies of updated grades, since the timing of these updates (and the original grades) comes into play with time sensitive grade models. If only *unique* grades were counted, there would be 7,667,032 in total, an 89.8 grades-per-student average to start and a 90.4 average at the end.

LMS grades are assigned throughout the semester, with earlier grades providing more utility for improving student outcomes while there still might be time to support strug-gling students. Within each timeframe in the right three columns of Table 6, we eliminated duplicate grades for a given grade item, keeping the last awarded grade. Figure 4 depicts the volume of grades-per-student entered within four-week timeframes. The proportion of grades assigned early in the semester has increased over time, supporting our earlier hypothesis about change in faculty usage. Specifically, the first two terms average 12.2 grades-per-student in the first half of the term while the last terms averaged 20.8, a 36% increase.

**Table 6.** Semester totals for student, courses, and LMS grades over nine semesters

|  | Students | Courses | LMS Grades | Unique Grades by Period: | | |
|---|---|---|---|---|---|---|
|  |  |  |  | First 4 Weeks | Next 4 Weeks | Rest of Term |
| Fall 2020 | 10,898 | 2,742 | 995,813 | 167,679 | 244,042 | 578,530 |
| Spring 2021 | 9,784 | 2,880 | 871,157 | 106,057 | 154,872 | 607,336 |
| Fall 2021 | 10,216 | 2,626 | 884,789 | 158,707 | 220,272 | 503,869 |
| Spring 2022 | 8,971 | 2,683 | 1,164,765 | 177,163 | 188,921 | 448,867 |
| Fall 2022 | 9,742 | 2,562 | 1,262,091 | 182,727 | 208,561 | 447,711 |
| Spring 2023 | 8,614 | 2,481 | 1,159,832 | 171,877 | 168,436 | 418,519 |
| Fall 2023 | 9,457 | 2,505 | 1,314,557 | 198,004 | 220,073 | 453,159 |
| Spring 2024 | 8,374 | 2,429 | 1,185,588 | 178,221 | 165,364 | 439,126 |
| Fall 2024 | 9,792 | 2,563 | 1,295,260 | 199,870 | 221,158 | 438,091 |
| Total | 85,848 | 23,471 | 10,133,852 | 1,540,305 | 1,791,699 | 4,336,397 |

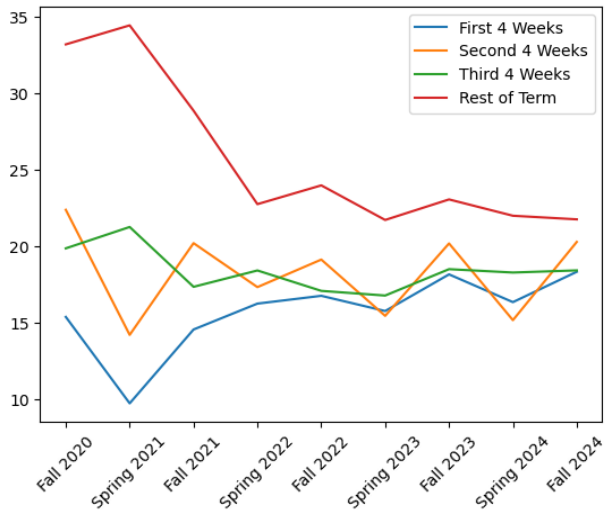

**Figure 4.** Unique grades per student across terms, disaggregated by the four-week period in which they were posted

Grades were assigned in a variety of ways. Our LMS allows for a "Points Numerator", a "Points Denominator", and / or a "Grade Value" for each grade. "Points Numerator" had 86,940 missing values, "Points Denominator" had 62,299 missing values, and "Grade Value" had 3,871 missing values and 1,364,804 non-float values. Hence, data analysis considered each of these as a predictor.

Since the numerator and denominator version of each grade has most numeric values (and an inherent weighting of grades given by denominator) we examined these by summing numerators, summing denominators, and dividing for each student (without regard to course). This minimized missing data since a student had a grade average if *any* of their courses had grades. For the remaining missing grades, we used median imputation, as we will throughout this analysis. We then employed machine learning to predict student semester GPA base *only* on LMS grades. We utilized flaml, which attempted to fit six models. ExtraTrees achieved the best prediction of semester GPA with an MSE of 0.455, 62% of the test set GPAs predicted within 0.5 grade point, and 85% predicted within 1 grade point.

On the other hand, converting the top 8 non-float Grade Values to floats (such as 95 for "A"), averaging all grades for each student, then executing machine learning also yielded best results using LGBM with an MSE of 0.385, 67% of the test set GPAs within 0.5 grade point, and 90% within 1 grade point. This is despite the fact that not every course has *any* LMS grades entered. So far, we are including even late semester grades which have limited utility in helping students change course. (This is the rightmost column of Table 7.) For comparison, the performance, as measured by $1-R^2$, was LGBM (0.3995), then RF (0.4001), then XGB_LimitDepth (0.4025), then XGBoost (0.4030), then ExtraTrees (0.4044).

Next, we recalculated both numerator-over-denominator and grade-average, first averaging all graded courses for each individual, then by weighting the average of those outcomes by the credit hours of each course. For the weighted numerator-over-denominator, LGBM performed best with an MSE of 0.406, 66% of the test set accurate to within 0.5 grade point, and 89% accurate to within 1 grade point. For the weighted grade-average, the best performance had an MSE of 0.420, 65% of the test set accurate to within 0.5 grade point, and 89% accurate to within 1 grade point.

Finally, considering the fact that at times optional assignments have points denominators (which would previously have effectively been assumed to be 0 if not completed), we calculated a median number of points for each course. We then divided each student-course numerator by the course median. We averaged across courses for each student, then calculated a weighted average across courses for each student. For the unweighted numerator-over-median, the best performing model, LGBM, had an MSE of 0.428, 61% of the test set accurate to within 0.5 grade point, and 89% accurate to within 1 grade point. For the weighted grade-average, now XGBoost performed best with an MSE of 0.398, 66% of the test set accurate to within 0.5 grade point, and 89% accurate to within 1 grade point.

Our perhaps unintuitive conclusion is that weighting grades by the denominator (the points declared possible by the instructor) and weighting grades in classes by the credit hours in that class both actually *decrease* the predictive power of grade data in this dataset. We restrict our attention to only the best feature.

We proceed to examine partial term data, since educational institutions want to know which students are predicted not be successful as early in the term as possible. We will restrict our attention to unweighted grade value data (with the most frequent string values converted to numeric, ex. A becomes 95). As seen in Table 7, grade data within the first 4 weeks allows for 80% accurate GPA prediction within 1 grade point, raising to 90% accuracy by the end of the term.

**Table 7.** The performance of grade data from different portions of the semester in predicting semester GPA

|  | First 4 Weeks | First 8 Weeks | First 12 Weeks | Full Semester |
|---|---|---|---|---|
| # of grades | 1,832,999 | 4,275,656 | 6,797,565 | 10,133,852 |
| MSE | 0.762 | 0.598 | 0.491 | 0.385 |
| within 0.5 point | 48% | 57% | 62% | 67% |
| within 1 point | 80% | 84% | 87% | 90% |
| Best model | LGBM | LGBM | LMGM | LMGM |

These early semester predictive findings are surprisingly good and seem to contradict Marques and colleagues' conclusion that LMS data alone was inadequate to meaningfully predict student success (2017). It is probable that LMS usage on university campuses has changed dramatically since 2017. More comparative work would be necessary to determine whether that was the primary difference in findings. Of critical importance, however, is the fact that 4 weeks of LMS grade data were sufficient to estimate 80% of students' semester GPA within one grade point.

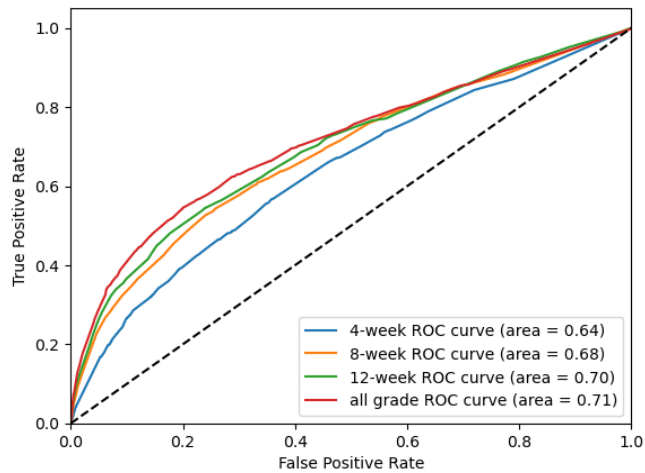

**Figure 5.** ROC Curves for predicting student discontinuance based on varying subsets of LMS grade data

Discontinuing college is more challenging to predict. Using the 4-week, 8-weeks, 12-week, and full grade data we also trained models with discontinuance as the outcome.

Figure 5 outlines the different ROC curves yielded from the 4 different data levels. Note that the area under curve (AUC) increases with each additional influx of data, but the increase is not dramatic. Best performance, as measured by 1-AUC, was achieved by XGBoost and XGB_LimitDepth (both 0.2880), followed by LGBM (0.2882), then ExtraTrees (0.2888), then RF (0.2896), then SGD (0.4968).

**Integrating LMS Data with Traditional Student Predictors**

By far the most common approach with LMS data is to combine it with other data sources to predict student performance. For instance, Bird and colleagues incorporated 250 course-specific predictors (in press).

In this analysis, we focus on midsemester analysis and use 20 factors (which produces a 1-2% improvement over 10 factors). The semester GPA prediction that performed best was the XGB_LimitDepth regressor as seen in Table 8. It achieved an MSE of 0.315, an $R^2$ of 0.6823, predicted 73% of the test set within 0.5 grade point of their GPA, and predicted 92% within 1 point of their GPA. Performance of the next best model, XGBoost, was about 1.5% worse using $1 - R^2$. Optimal configurations for each model can be found in the Appendix.

**Table 8.** Performance of various models predicting the 3 key outcomes using 20 features

|  | Semester GPA ($1-R^2$) | Overall GPA ($1-R^2$) | Discontinuance (1-AUC) |
|---|---|---|---|
| LGBM | 0.3231 | 0.1482 | 0.1433 |
| Random Forest | 0.3321 | 0.1641 | 0.1505 |
| XGBoost | 0.3226 | 0.1477 | 0.1543 |
| ExtraTrees | 0.3299 | 0.1531 | 0.1499 |
| XGB_LimitDepth | **0.3177** | **0.1451** | **0.1428** |
| SGD | 0.7532 | 0.7933 | 0.3303 |
| LRL | 1 |  | 0.2455 |

As seen in Table 8, predicting overall GPA was also modeled best using the XGB_LimitDepth regressor, which achieved an MSE of 0.095, an $R^2$ of 0.8549, predicted 93% of the test set within 0.5 grade point of their GPA, and predicted 98% within 1 point of their GPA. The performance of the next best model, XGBoost, was 1.8% worse.

Finally, the best model found for discontinuance was XGB_LimitDepth classifier. It achieved accuracy of 0.91 and had an area under ROC of 0.8572. The second-best model, LGBM, was only 0.35% behind as seen in Table 8.

We calculated SHAP values on all three models for the approximately 17,000 values in the test set. As seen in Figure 6, the mean influence of LMS logins and grades on the test set was substantial – the 2$^{nd}$ and 4$^{th}$ most significant features for semester GPA, the 3$^{th}$ and 5$^{th}$ for overall GPA, and 2$^{nd}$ and 6$^{th}$ for discontinuance.

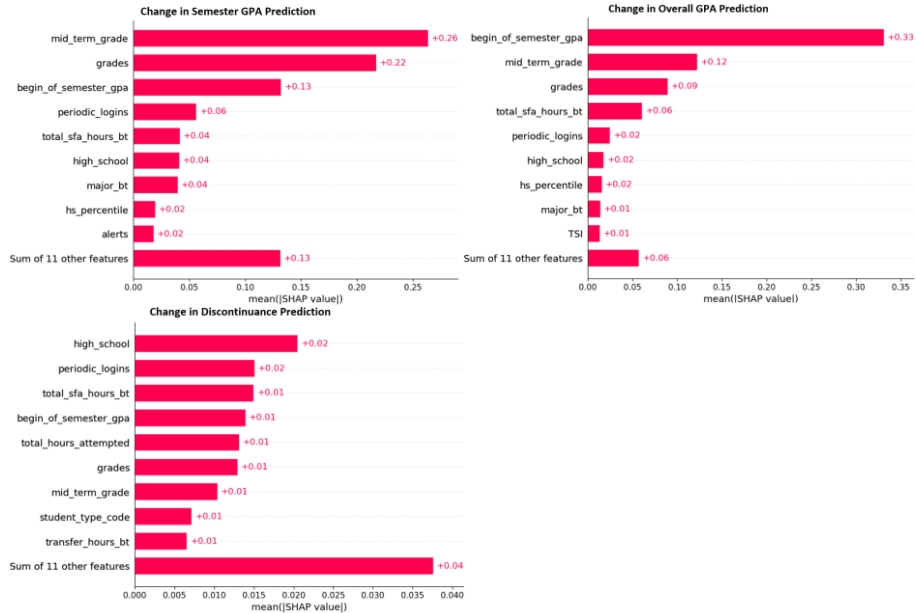

**Figure 6.** Mean absolute SHAP values of features for 3 key outcomes

## 4    Applications

These predictive models are of great interest to university stakeholders on our campus. Since these models can be informed by midsemester LMS data, the primary model in practice, a semester GPA predictor, is updated every Monday and provided to academic advisors that serve approximately 60% of undergraduate students. Another cabinet-level administrator is also interested in leveraging these predictive models to inform Student Affairs interventions. They hope to make use of the overall GPA model, but may also use the discontinuance model.

Currently, raw counts of LMS logins, tutor visits, and other dynamic features are provided to advisors to help them interpret the model's predictions. However, the authors are exploring the deployment of SHAP values to assist student support staff in interpreting model predictions for specific students.

Finally, there appears to be tremendous interest in models for specific populations for which (a) predictions are particularly accurate such as full-time students, Psychology students, or sophomores and juniors (as illustrated in Figures 2 and 3); or (b) specific interventions might be particularly useful (such as career coaching for students not on track to be accepted into the nursing program). Work on specialized models for sub-populations is ongoing.

## 5    Conclusion

LMS data provides vital insight into the academic well-being of students midsemester. The quality of the analysis of these data can dramatically increase their predictive power, as seen especially with logins versus periodic logins. Further, LMS usage has changed in our context, even since 2021. Accounting for differences between legacy data will be particularly relevant for LMS data. The intention of this article is not to *tell* other institutions which format of their data will be most predictive, but to *encourage* other institutions to carefully examine their unique data characteristics and to provide ideas about key aspects to analyze.

Different engineered features appear to have different strengths, but in the present dataset it appears that attempts to weight LMS grades are counterproductive. Grades across classes seem to do a good job of predicting overall GPA performance, even if some classes are not represented by grades in the LMS. LMS logins, however, appear to do a better job of predicting discontinuance from college.

LMS data is a vital tool in improving overall predictive models for student success, but may be among the most sensitive features because of their volume, the variety of who entered them, and their multifaceted interpretations.  In the future, it is likely that learning management systems will be used even more heavily in education. The data science community would be wise to continue to study and share insights on this important topic.

**Acknowledgments.** This study was made possible in part by a faculty development leave award from Stephen F. Austin State University.

**Disclosure of Interests.** The authors have no competing interests to declare that are relevant to the content of this article.

## Appendix: Best Configurations of Estimators Considered

**Prediction 1: semester_gpa**

Features: mid_term_grade, grades, begin_of_semester_gpa, alerts, high_school, periodic_logins, hs_percentile, major_bt, content_completion_rate, content_topic_visits, TSI, college_bt, new_SAT, admit_code , total_time, transfer_hours_bt, total_sfa_hours_bt, student_type_code, meal_plan, hall_name

LGBM best configuration: {'n_estimators': 1173, 'num_leaves': 10, 'min_child_samples': 6, 'learning_rate': 0.054987380085078814, 'log_max_bin': 9, 'colsample_bytree': 0.885652480919423, 'reg_alpha': 0.009984034615570498, 'reg_lambda': 0.2388700003862822}

RF best configuration: {'n_estimators': 63, 'max_features': 0.6739758718573503, 'max_leaves': 1057}

XGBoost best configuration: {'n_estimators': 442, 'max_leaves': 360, 'min_child_weight': 20.751061790819115, 'learning_rate': 0.03399093100192437, 'subsample': 0.8809985595678514, 'colsample_bylevel': 0.8490633393816673, 'colsample_bytree': 0.7514653053266084, 'reg_alpha': 0.006254798766353442, 'reg_lambda': 0.28534114950225703}

ExtraTrees best configuration: {'n_estimators': 71, 'max_features': 0.8316456275359224, 'max_leaves': 4760}

XGB_LimitDepth best configuration: {'n_estimators': 2199, 'max_depth': 10, 'min_child_weight':6.137984684618797, 'learning_rate': 0.007759642790692926, 'subsample': 1.0, 'colsample_bylevel': 0.7316290044540736, 'colsample_bytree': 0.8903447372116292, 'reg_alpha': 0.0010515900751535718, 'reg_lambda': 4.050395673709929}

SGD best configuration: {'penalty': 'l2', 'alpha': 2.073827860066681e-05, 'l1_ratio': 0.9999999999999999, 'epsilon': 0.04721809031174382, 'learning_rate': 'optimal', 'eta0': 0.013646120811719356, 'power_t': 0.27563902198714846, 'average': False, 'loss': 'epsilon_insensitive'}

**Prediction 2: end_of_term_gpa**

Features: begin_of_semester_gpa, mid_term_grade, grades, hs_percentile, high_school, alerts, major_bt, periodic_logins, content_completion_rate, content_topic_visits, TSI, new_SAT, college_bt, admit_code, student_type_code, transfer_hours_bt , total_time, total_sfa_hours_bt, meal_plan, hall_name

LGBM best configuration: {'n_estimators': 557, 'num_leaves': 17, 'min_child_samples': 15, 'learning_rate': 0.042242947825983146, 'log_max_bin': 9, 'colsample_bytree': 0.5623352336965426, 'reg_alpha': 0.0009765625, 'reg_lambda': 0.07755818552475893}

RF best configuration: {'n_estimators': 162, 'max_features': 0.584414084625783, 'max_leaves': 2263}

XGBoost best configuration: {'n_estimators': 950, 'max_leaves': 60, 'min_child_weight': 7.233785289465286, 'learning_rate': 0.015126332186899651, 'subsample': 1.0,

'colsample_bylevel': 0.3685988387608087, 'colsample_bytree': 0.8129075176371072, 'reg_alpha': 0.0019706937086253853, 'reg_lambda': 0.1148611342151243}

ExtraTrees best configuration: {'n_estimators': 250, 'max_features': 0.7088416535926986, 'max_leaves': 12114}

XGB_LimitDepth best configuration: {'n_estimators': 228, 'max_depth': 13, 'min_child_weight': 40.26812114743653, 'learning_rate': 0.06219377762209235, 'subsample': 0.8365749881305449, 'colsample_bylevel': 0.8488288054460789, 'colsample_bytree': 0.785498557164527, 'reg_alpha': 0.09749190094363033, 'reg_lambda': 47.50558410170101}

SGD best configuration: {'penalty': 'l2', 'alpha': 3.57153377370962e-05, 'l1_ratio': 0.9999999999999999, 'epsilon': 0.006407201316208552, 'learning_rate': 'optimal', 'eta0': 0.008615774063807258, 'power_t': 0.3052063818916277, 'average': False, 'loss': 'epsilon_insensitive'}

**Prediction 3: discontinued**

Features: high_school, periodic_logins, student_type_code, grades, major_bt, total_hours_attempted, TSI, mid_term_grade, begin_of_semester_gpa, admit_code, total_sfa_hours_bt, college_bt, content_topic_visits, alerts, content_completion_rate, transfer_hours_bt, hs_percentile, total_meals, first_gen_status, meal_plan

LGBM best configuration: {'n_estimators': 2541, 'num_leaves': 1667, 'min_child_samples': 29, 'learning_rate': 0.0016660662914022304, 'log_max_bin': 8, 'colsample_bytree': 0.5157078343718623, 'reg_alpha': 0.045792841240713165, 'reg_lambda': 0.0012362651138125363}

RF best configuration: {'n_estimators': 1000, 'max_features': 0.1779692423238241, 'max_leaves': 7499, 'criterion': 'gini'}

XGBoost best configuration: {'n_estimators': 13499, 'max_leaves': 60, 'min_child_weight': 0.008494221584011285, 'learning_rate': 0.006955765856675575, 'subsample': 0.5965241023754743, 'colsample_bylevel': 0.590641168068946, 'colsample_bytree': 1.0, 'reg_alpha': 0.2522240954379289, 'reg_lambda': 5.351809144038808}

ExtraTrees best configuration: {'n_estimators': 2047, 'max_features': 0.46132798093546956, 'max_leaves': 12856, 'criterion': 'gini'}

XGB_LimitDepth best configuration: {'n_estimators': 877, 'max_depth': 11, 'min_child_weight': 0.6205465771093738, 'learning_rate': 0.013622118381700795, 'subsample': 0.566692814245426, 'colsample_bylevel': 0.8865741642101924, 'colsample_bytree': 1.0, 'reg_alpha': 0.01386336444764391, 'reg_lambda': 3.113947886074155}

SGD best configuration: {'penalty': 'l2', 'alpha': 0.0001, 'l1_ratio': 0.1500000000000002, 'epsilon': 0.1, 'learning_rate': 'invscaling', 'eta0': 0.010000000000000005,

'power_t': 0.5, 'average': False, 'loss': 'modified_huber'}

LRL 1 best configuration: {'C': 1.0}